# Effect of Franchised Business models on Fast-Food Company Stock Prices in Recession and Recovery – Weibull Analysis

*Sandip Dutta[1] and Vignesh Prabhu*
Clemson University
Clemson, SC
[1]Contact Author: sdutta@clemson.edu

Abstract:

At the initial stages of this research, the assumption was that the franchised businesses perhaps should not be affected much by recession as there are multiple cash pools available inherent to the franchised business model. However, after analyzing the available data, it indicated otherwise- the stock price performance as discussed indicates a different pattern. The stock price data is analyzed with an unconventional tool – Weibull distribution and observations confirmed the presence of either a reverse trend in franchised business than what is observed for non-franchised or the franchised stock followed large food suppliers. There is a layered ownership and cash flow in a franchised business model. The parent company run by franchiser depends on the performance of child companies run by franchisees. Both parent and child companies are run as independent businesses but only the parent company is listed as a stock ticker in stock exchange. Does this double layer of vertical operation, cash reserve, and cash flow protect them better in recession? The data analyzed in this paper indicates that the recession effect can be more severe; and if it dives with the average market, expect a slower recovery of stock prices in a franchised business model. This paper characterizes the differences and explains the natural experiment with available financial data.

Prelude:

Please note that this is a work in progress. Student author, Mr. Prabhu, will be graduating soon and therefore will not have time to produce better quality plots. Expecting a slow progress, we decided to publish this work in progress paper as it does contain important thought-provoking information. Please send us any comments and questions you may have to the primary contact: Dr. Sandip Dutta, sdutta@clemson.edu. We hope to expand this work to a full-blown high-quality paper to be very useful for the community, and your help is appreciated.

Introduction:

Since this paper discusses franchised business model and compare the stock price movement with non-franchised businesses, we take this opportunity to define terms and concepts used in this paper to better acquaint the reader with the discussions. We discuss the Weibull and related probability distribution on the stock price movement of franchised and non-franchised fast food businesses. The discussion focuses on the franchising, earnings management, franchise fairness, restaurant, analyses of spillover from the recession, and some aspects of Weibull distribution.

There are several publications available on franchising and Blair and Lafontaine (2005) explained the underlying business structure in a franchised business quite thoroughly. They showed vertical price control in franchising and explained the vertical constraints of franchising. A researcher interested in more details is encouraged to read their publications. Among other notable publicly published articles- Franchising 101 (1998) explains the structure of franchised business, related documents, and legal structures that bind franchised units together.

Caves and Murphy (1976) noted that Franchised businesses accounted for over 38 percent of all retail sales in the United States in mid-seventies; and those franchised businesses originated 12 percent of the gross national product. This referenced paper used contemporary developments in the modeling of firms and market structures to explain the occurrence of franchising and its distribution among different sectors of the economy. They analyzed the contractual consequences of profit maximization by the franchisor.

Martin (1988) observed that Franchising is an important but controversial form of vertical integration. Allegations of opportunistic behavior by franchisors have led to calls for public regulation and, in some states, "fairness in franchising" laws. The advisability of such regulation depends on the long-run incentives to franchise. This referenced paper offered evidence on the incentives to franchise at that time.

Felstead (1993) explained how and why franchising works. He also made observations with data to illustrate how franchising process was binding firms together. The essential concept of franchising is well defined in Wikipaedia: "Franchising is based on a marketing concept which can be adopted by an organization as a strategy for business expansion. Where implemented, a franchisor licenses its know-how, procedures, intellectual property, use of its business model, brand, and rights to sell its branded products and services to a franchisee. In return the franchisee pays certain fees and agrees to comply with certain obligations, typically set out in a Franchise Agreement."

We select franchised businesses related to the fast food industry as they are better understood than other complicated business models like educational services. Farfan (2019) observed that even if visually chain restaurant stores occupy the populated landscape, in reality only about 30 percent of America's restaurants are part of a multi-unit chain, and only a fraction of those restaurant chains are publicly traded restaurant companies. That is why we could not get too many data options. Fortunately, most of the largest multi-unit restaurant companies in the U.S. are publicly traded on U.S. stock exchanges and not privately owned.

Management of earnings and accounting manipulations in franchising are discussed by Gim et al. (2019). According to them, earnings management or manipulations have become popular with many firms. This makes it difficult to make decisions based on the financial statements. The financial accounting literature has made efforts to identify the determinants of earnings management behavior in various industries. However, for the restaurant industry it was found to be necessary to consider the effects of franchising to accurately understand earnings management behaviors because franchise restaurants vastly differ from non-franchise restaurants in terms of raising capital and information asymmetry. Therefore, the main purpose of their study was to investigate whether franchising as a firm characteristic caused any meaningful differences in the earnings management behavior of the restaurant firms. They showed that during the growth phase (survival mode), franchise restaurants engaged more actively in earnings management than non-franchise restaurants. Further, the deterrence effect of financial leverage on earnings management was weaker for franchise restaurants than non-franchise restaurants. Sometimes cash flow between franchisee and franchisor were published as positive cash flow for the public company, even if it was internal cash flow. Overall, their work suggested that franchise restaurants were generally more inclined towards earnings management. Since current work compares franchised data with non-franchised stores, this accounting manipulations may hide some important information that cannot be found without forensic accounting.

The scope of our paper is limited to the public companies listed in the US stock exchange, but most of these businesses operate worldwide and the rules and regulations vary from country to country. There are some publications that address international franchising, but we will not discuss too much in that front. The franchising structure of European banks are discussed by Jonghe and Vennet (2008) and interested reader can get references in European businesses from them. This referenced paper investigated how stock market investors perceive the impact of market structure and efficiency on the long-run performance potential of European banks. Authors used a modified Tobin's Q ratio as a measure of bank franchise value. This measure was applied to discriminate between the market structure and efficient-structure hypotheses in a coherent forward-looking framework. Their results showed that banks with better management or production technologies possessed a long-run competitive advantage. In addition, bank market concentration did not affect all banks equally.

Sayabaev et al. (2016) discussed franchising effect on tourism in Kazakhstan. According to them, the tourist business was an attempt of adaptation for the country in international arena. An improvement of tourism increased the credibility of business entities in the country. They investigated the effect of franchising on local tourism and perhaps that can be extended to similar economic scenarios. The referenced article is useful for entrepreneurs, analysts, teachers and students. They discussed in detail the development history of the franchise business in Kazakhstan. They presented a detailed chronology of major development, pointed the scope of the franchising as well as domestic companies using franchising to grow. They determined the importance of franchising in the development of small and medium business in Kazakhstan. Theoretical and methodological basis of research served as the scientific concept of modern economic theory, theories of innovation, and theories of management; and also provided  new research avenues for domestic and foreign scholars. Authors came to the conclusion that franchising has developed international business collaborations with local business developments. They also observed franchising brought in additional capital and technical know how to improve the local economy.

Gim and Jang (2019) examined industry-specific factors in order to explain the dividend behaviors of franchise and non-franchise restaurants. Due to the unique differences in characteristics between them, they argued - it may not be appropriate to view the restaurant industry as one entity displaying homogenous dividend behaviors. Their study demonstrated that the dividend behaviors of franchise restaurants were a function of the size of institutional holdings and prior dividend payments, while those of non-franchise restaurants depended on the degree of growth opportunity and financial leverage. The findings further implied that franchise restaurants with higher free cash flows and the resulting information asymmetry closely followed agency theory and signaling theory, whereas non-franchise restaurants with insufficient internal cash strictly followed the pecking order theory.

Rhoua et al. (2019), argued that a firm's risk-taking behavior is linked with managerial ownership and firm performance. Managerial ownership affects corporate risk-taking in industries characterized by high financial and operational risks, like the restaurant industry. To understand this important and understudied link, they considered agency theory to examine the relationship between managerial ownership and franchising; that was typically used as a risk-reduction strategy of restaurant firms. Their results from panel data analyses using a sample of 962 firm-year observations showed that managerial ownership was negatively associated with degree of franchising. Further, they found that after considering the scope of managerial discretion, there was a U-shaped relationship between managerial ownership and degree of franchising such that the degree of franchising decreased as managerial ownership increased up to a certain level, but then increased in tandem as managerial ownership increased to higher levels. That indicated, there was an optimal level of franchising associated with managerial ownership, implying that franchisers could influence their firms' risk-taking behavior by setting target managerial ownership goals and designing effective incentive contracts.

In recent years, mobile food business is getting popular and some are using franchising to fuel their growth. However, these are mostly small operations and not much information is available from the stock market. Jennings (2009) observed, the mobile vending craze is switching into a new gear with franchising. Author observed, as consumers continue their infatuation with gourmet fare on wheels, some mobile vendors are working to roll their success forward by replicating their trucks through licensing and franchise deals.

Large franchisors have a mix of franchised and corporate owned businesses. Hsu and Jang (2009) found that Franchising has significantly affected the US economy by contributing to a rapid growth of its retail sales. To identify whether franchising influenced a restaurant firm's financial performance, their study investigated the profitability and intangible values of both franchised and non-franchised restaurant firms; and the effect of the combination of franchised and company-owned outlets of restaurant firms (i.e., franchise proportion). The results of their study showed that franchised firms had significantly higher profitability than non-franchised firms; and the relationships between franchise proportion and firm profitability and intangible value were curvilinear (inverted U-shape), verifying the existence of an optimal franchise proportion. They proposed that restaurant franchisors could maximize their profitability and intangible values with an optimal franchise proportion when other variables were held constant, implying that it was important to pay attention to the franchise proportion together with other management strategies.

Salar and Salar( 2014 ) prepared an apprentice's perspective on the franchised business. Their short report was an attempt to compare the advantages and disadvantages of franchising as a newcomer's perspective. Using SWOT analysis, franchising was explained with brand recognition, low risk to failure, easy setup, ready customer portfolio and easy to find financial support. They also showed the disadvantages of franchising as strict rules, dependency and high cost. After balancing the advantages over disadvantages of franchising it was concluded that the franchising was more advantageous. However, there is room to argue on their superficial observations, but it should be noted that this article captures the essence of franchising perspective that is presented to newcomers.

Solı́s-Rodrı́guez and Gonzá́lez-Dı́az (2017) identified differences in contract design between successful and less successful franchise chains. Comparing contracts from both groups of companies, they found that franchise contracts were unbalanced irrespective of the chain's success: contracts covered more of franchisees' obligations than franchisors' obligations. They also observed that contracts in successful franchise chains were more complete (i.e. it covered a larger number of contingencies) than the less successful ones; and this was observed in the contingencies regarding franchisees' obligations, which were more fully covered in the contracts of more successful chains. More specifically, within the contingencies regarding franchisees' obligations, successful chains restricted the franchisee decision rights more frequently on day-to-day business operations than on financial conditions or post-contractual contingencies. Successful chains were more sensitive to franchisees' opportunistic behavior, because they learnt how to manage and solve any potential conflicts, or maybe it was because of differences in bargaining power. Their study also found, franchisors' obligations were not statistically different between groups, which they interpreted as evidence that relational contracting mechanisms did not substitute formalization.

Combs et al., (2004) explained franchising based on resource scarcity and agency theory, with the empirical findings regarding three key franchising constructs—franchise initiation, subsequent propensity to franchise, and franchise performance. They suggested that research emphasis needed to shift toward understanding why firms initiate franchising and how franchising impacted different types of organizational performance. They also found that current research could benefit from additional theoretical diversity and thus they offered new propositions grounded in three theories that were not widely applied to franchising.

Wilson and Shailer (2015) examined dual distribution system that used both franchisor-operated and franchisee-operated outlets. Their work implied, a franchisor's information disadvantage was reduced when contracting with franchisee retailers. Using detailed qualitative and quantitative managerial data, they found persuasive evidence of the strategic use of performance information obtained from franchisor operated outlets to reduce information asymmetry and to enhance contracting efficiency for franchisee-operated outlets. They tested whether the proximity of franchisor-operated retail outlets to franchisee-operated retail outlets reduced underpricing of quasi-franchise contracts. Their results supported the proposition that information asymmetry reduced contracting efficiency. Their findings indicated that a manufacturer can reduce intrinsic information asymmetry by maintaining franchisor-operated outlets that were geographically proximate to the franchisee-operated outlets; and that improved the franchisor's pricing of franchising contracts. They concluded that dual distribution reduced the franchisor's information asymmetry and increased their contract pricing efficiency.

The spillover effect of US recession in 2008-2009 to other stock markets around the world were felt and were studies extensively. The spillover study techniques are relevant to this work and studies related to spillover to China, Hong Kong, and emerging markets are discussed here briefly to build an understanding on the connectivity among world markets. These studies also helped to identify any disconnect in the franchised company stock from the overall stock market. Frank and Hesse (2009) studied the spillover of US financial crisis to emerging markets. They developed mathematical models to correlate the market trends. Their main findings suggested that implied correlations between the U.S. Libor-OIS spread, a proxy for funding illiquidity, and EMBI+ sovereign bond spreads of Asia, Europe, and Latin American countries, sharply increased following the onset of the subprime crisis. In addition, the Shanghai stock market correction in February 2007 led to a temporary spike of the correlation measures, whereas the Lehman collapse caused the largest increase of co-movements among these variables. Similarly, the relationship between the S&P 500 and the EMBI+ regional bond spreads exhibited a potential break during the Chinese episode, after which correlations increased from the beginning of the subprime crisis and reached their peak after the Lehman failure.

Zhang and Sun (2009) analyzed spillovers of the United State's subprime financial turmoil to mainland China and Hong Kong (HK). Their paper examined whether the U.S. subprime financial turmoil had any statistically significant effect on both the daily returns and the conditional volatility of stock prices in China and HK. To capture the possible spillover effects, they employed a two-stage procedure; in the first stage univariate GARCH models were used, and in the second stage multivariate GARCH models were developed to further identify the sources and magnitudes of the spillovers. Using both univariate and multivariate GARCH models indicated China's stock market was not immune to the financial crisis. It was clearly evidenced by the price and volatility spillovers from the United States to China in MGARCH models. They also observed Hong Kong's equity returns exhibited more significant price and volatility spillovers from the United States, and past volatility shocks in the United States had a more persistent effect on future volatility in HK than in China, indicating that HK was more affected due to its role as an international financial center. Less cross-volatility between Hong Kong and China than between the United States and China showed that the impact of the United States on China was greater than that of Hong Kong in the context of volatility persistence, due mainly to the United States as the origin of the subprime crisis.

To close this discussion, we like to bring in some of the Weibull based studies applied on stock trading and event predictions. Kizilersu et al. (2016) studied consequences of the digital revolution on share trading. They researched trading in the London Stock Exchange and found that if ultra-high frequency manipulation on time scales less than around ten milliseconds is excluded, all relevant changes in the order book happened with time differences that were randomly distributed and well described by a left-truncated Weibull distribution with universal shape parameter (independent of time and same for all stocks). The universal shape parameter corresponded to maximum entropy (uncertainty) of the distribution. Unlike engineering, the entropy in financial markets means the uncertainty and volatility.

Weibull distribution can be used to predict the natural disaster patterns like flooding based on historical data. Unlike hurricane or wind based natural disasters, flooding is limited by the presence of a river and related river basin. So, it makes sense to characterize the flood patterns based on historical data. Bo et al. (2019) studied high water levels of the large rivers that can lead to flooding and loss of property for tens of thousands of residents. The information from the systematical monitoring of the water levels allowed them to obtain probability distributions for the extremely high values of the water levels of the rivers of interest. They studied time series containing information from more than 10 years of satellite observation of the water level of the Huang He river (Yellow river) in China. They showed that the corresponding Weibull distribution may help evaluating the risks associated with floods for the population and villages in the corresponding region of the Huang He river.

Use of Weibull or Gamma distribution in a stock market analysis is rare but not absent. Rocha et al. (2014) used available data from the New York stock market (NYSM) to fit four different distribution models with the correspondent volume-price distributions at each 10-minute lag. Models used were- the Gamma distribution, the inverse Gamma distribution, the Weibull distribution and the log-normal distribution. The volume-price data, which measures market capitalization, appears to follow a specific statistical pattern, other than the evolution of prices measured in similar studies. They found that the inverse Gamma model gave a superior fit to the volume-price evolution than the other models. The inverse Gamma distribution was applied on NYSM data and they analyzed the evolution of its distribution parameters as a stochastic process. Assuming that the evolution of those parameters was governed by coupled Langevin equations, they derived the corresponding drift and diffusion coefficients, which then provided insight for understanding the mechanisms underlying the evolution of the stock market.

Stock market traders are well connected and there are many mathematical models available to illustrate the links and consumer behavior on the fluctuations in stock price.

Network of trade analysis seems to have a logarithmic return pattern as observed by daCruz and Lind (2013). They considered the evolution of scale-free networks according to preferential attachment schemes and showed the conditions for which the exponent characterizing the degree distribution is bounded by upper and lower values. Their framework was an agent model, presented in the context of economic networks of trades, which showed the emergence of critical behavior.

Methodology and Data Source:

SAS software was used to analyze and develop the plots presented here. Table 1 shows the companies selected for this study- they were related to retail food and processed food distribution. MCD, WEN, and DPZ are well known franchised fast food retailers. SBUX, CBRL, CMG are large food retailers which are mostly not franchised. K, KO, and PM are large processed food manufacturers.

| Ticker | Company Name | Franchised |
|---|---|---|
| MCD | Mcdonald's Corp | Yes |
| WEN | Wendys Co | Yes |
| DPZ | Domino's Pizza, Inc. | Yes |
| SBUX | Starbucks Corporation | Mixed |
| CBRL | Cracker Barrel Old Country Store, Inc. | No |
| CMG | Chipotle Mexican Grill, Inc. | No |
| K | Kellogg Company | No |
| KO | Coca-Cola Co | No food outlet |
| PM | Philip Morris International Inc. | No |

Table 1: Companies selected for this study

| Ticker | Mu | Sigma |
|---|---|---|
| MCD | 4.012 | 0.096 |
| WEN | 2.006 | 0.562 |
| DPZ | 2.458 | 0.5301 |
| SBUX | 2.2177 | 0.3835 |
| CBRL | 3.4525 | 0.0105 |
| CMG | 4.3901 | 0.2931 |
| K | 3.9004 | 0.10006 |
| KO | 3.2538 | 0.1161 |
| PM | 3.8179 | 0.1296 |

Table 2: Distribution of stock price data as plotted in Figure 10.

Results and Discussions:

Results are presented in three sections. First part shows the raw stock prices of the companies listed in Table 1. Figure 1 shows the raw data of daily closing prices of these stocks. Figures 2-9 shows the annual performance data available from annual financial statements at https://www.macrotrends.net/. Figure 5, EPS plot needs to be adjusted as WEN had a negative earning and that made the plot not readable for WEN.

Figure 10 shows the statistical analysis of the stock prices using different probability distributions available in SAS. To create these plots, we took the daily closing prices as sample and processed the corresponding distribution. The normalized average and standard deviation are shown in Table 2. Note that this plot does not use time series data; instead we have used the closing prices without the time stamp to analyze these distributions. A time series data analysis is in process and will follow this publication. Table 2 shows that the franchised companies either show small variance or very high variance as compared to other non-franchised operations. The low variance results of MCD are surprisingly close to the average and variances observed in non-franchised food producers like K, Ko, and PM.

The franchised industry has a buffer layer between the parent franchisers and direct customers. Franchisees often have additional sources of cash of their own and other reserves like investments and home equity to sustain their businesses. This buffer layer creates a different stock price distribution in the daily variations, even if the financials do not show a marked difference among these franchised and non-franchised companies as shown in Figure 2-9. The franchised company like MCD (Fig. 1a) shows the resilience and holds the stock price steady even in the average market meltdown. Whereas WEN (Figure 1b) and DPZ (Figure 1c) fall below the sustainable level and cannot recover along with the overall market as is shown by other similar food related companies in Figure 1.

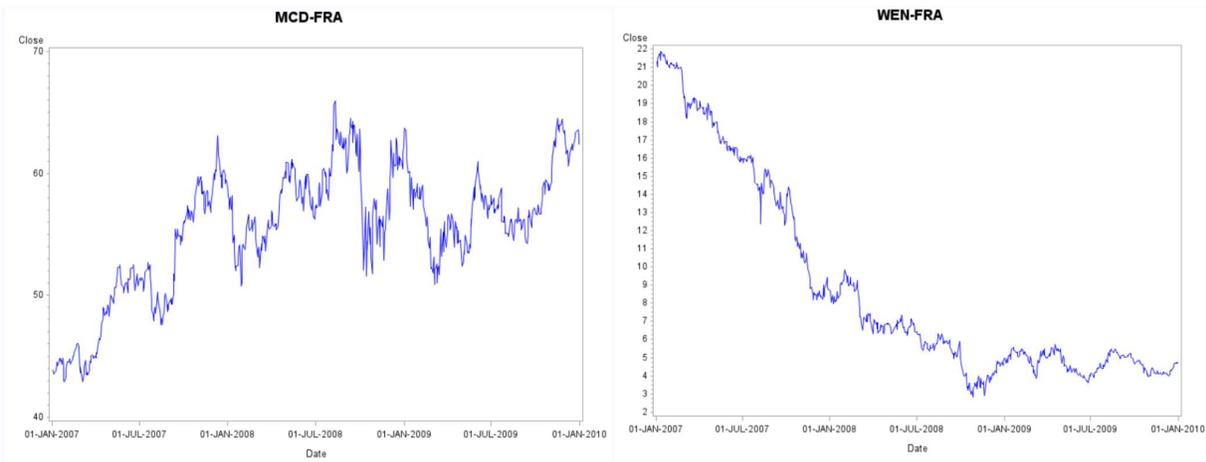

Fig 1a: Stock Price data MCD – McDonald's. Fig 1b: Stock Price data WEN – Wendy's

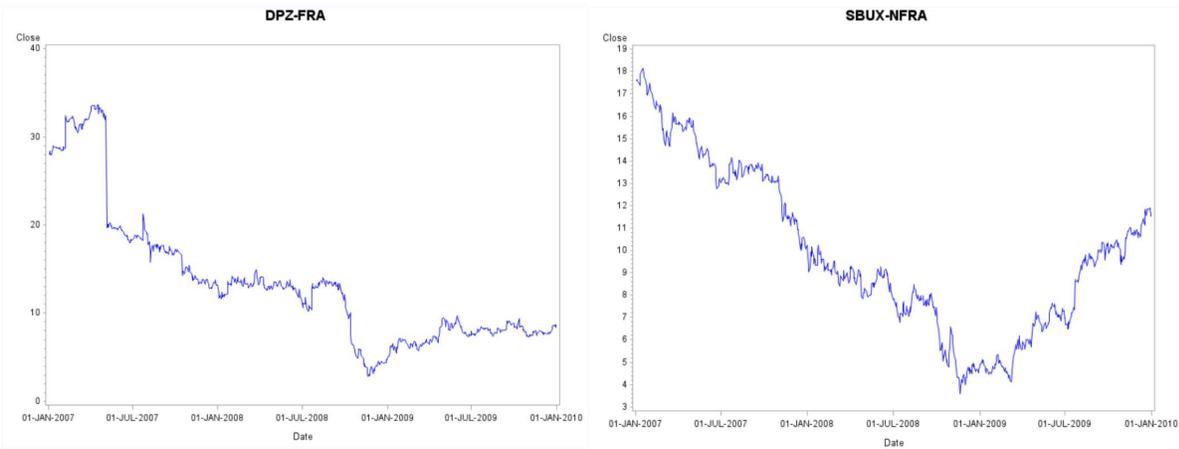

Fig 1c: Stock Price data DPZ – Domino's Pizza. Fig 1d: Stock Price data SBUX - Starbucks

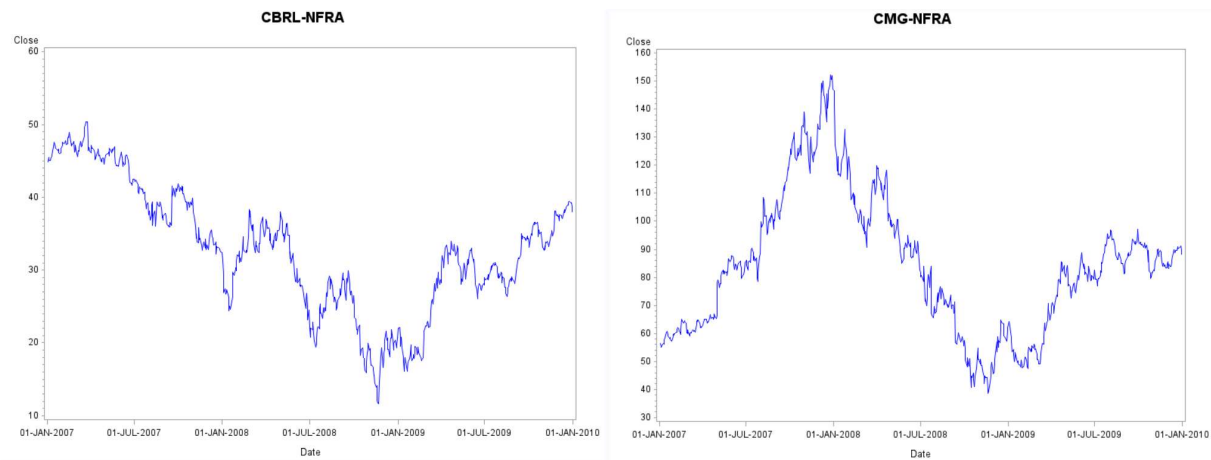

Fig 1e: Stock Price data CBRL- Cracker Barrel. Fig 1f: Stock Price data CMG - Chipotle

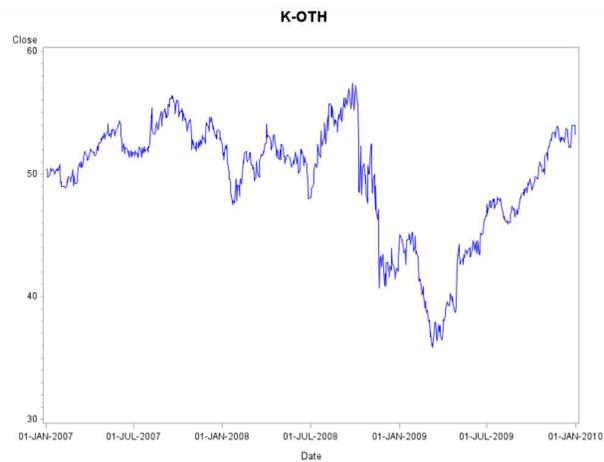 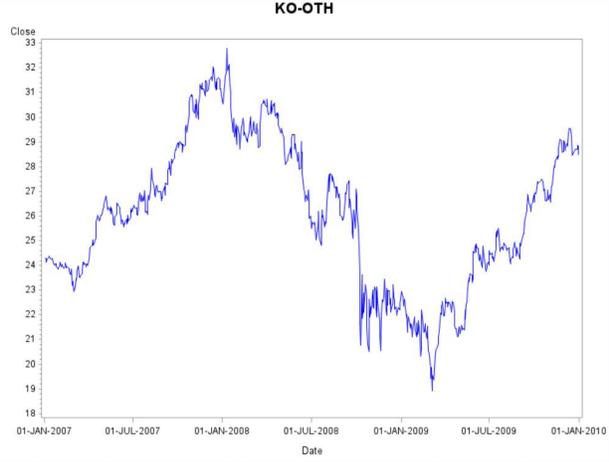

Fig 1g: Stock Price data K – Kellogg.   Fig 1h: Stock Price data KO – Coca Cola

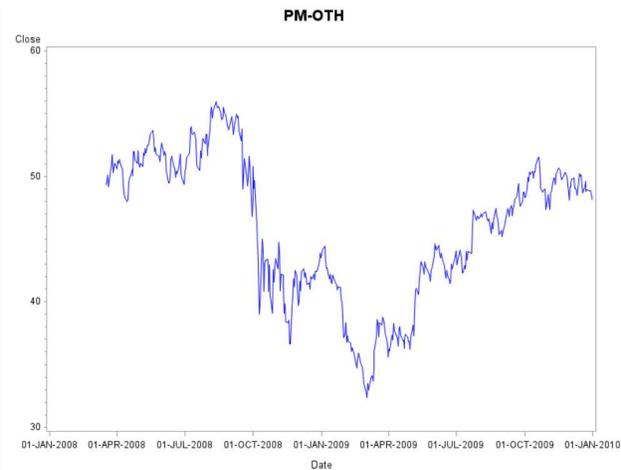

Fig 1i: Stock Price data PM – Phillip Morris

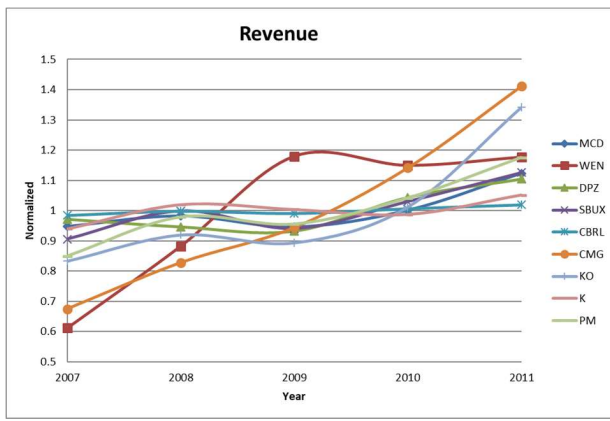 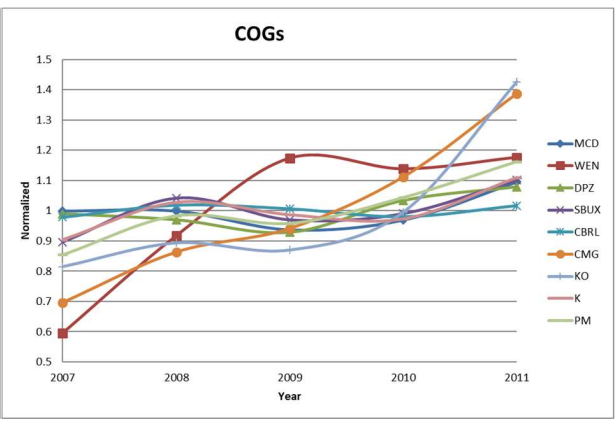

Fig 2: Revenue normalized over 2007-2011.      Fig 3: Cost of Goods normalized over 2007-2011

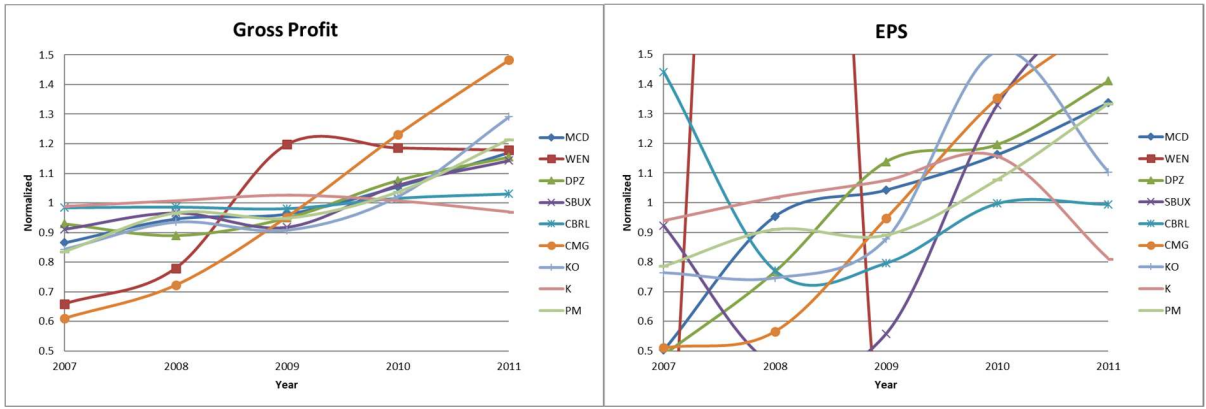

Fig 4: Gross Profit normalized over 2007-2011. Fig 5: Earnings per share normalized 2007-2011

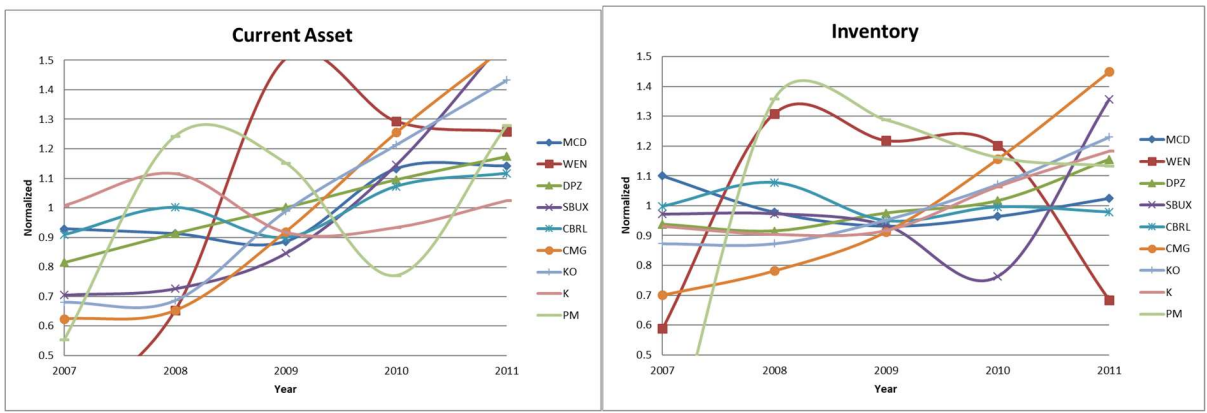

Fig 6: Current Asset normalized over 2007-2011. Fig 7: Inventory normalized over 2007-2011

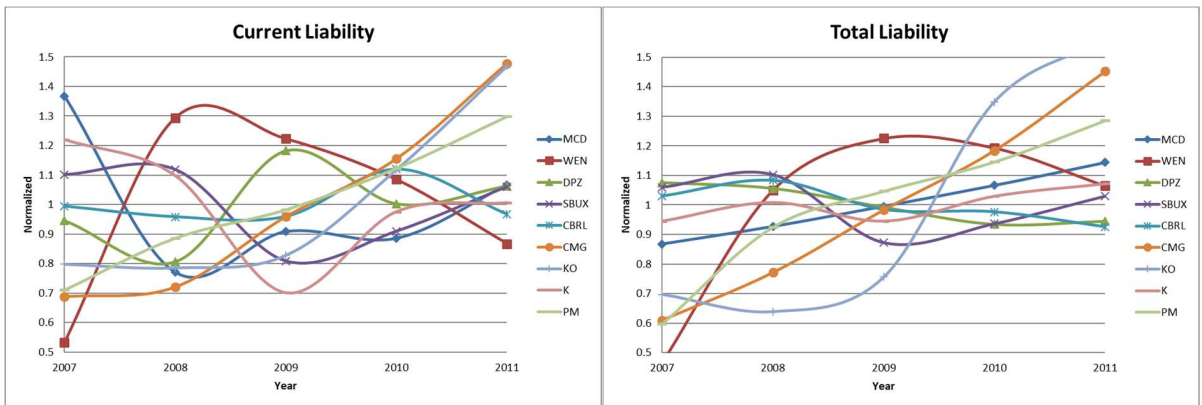

Fig 8: Current Liability normalized over 2007-2011. Fig 9: Total Liability normalized 2007-2011

Fig. 10: Details of Weibull and other related PDF plots are listed below

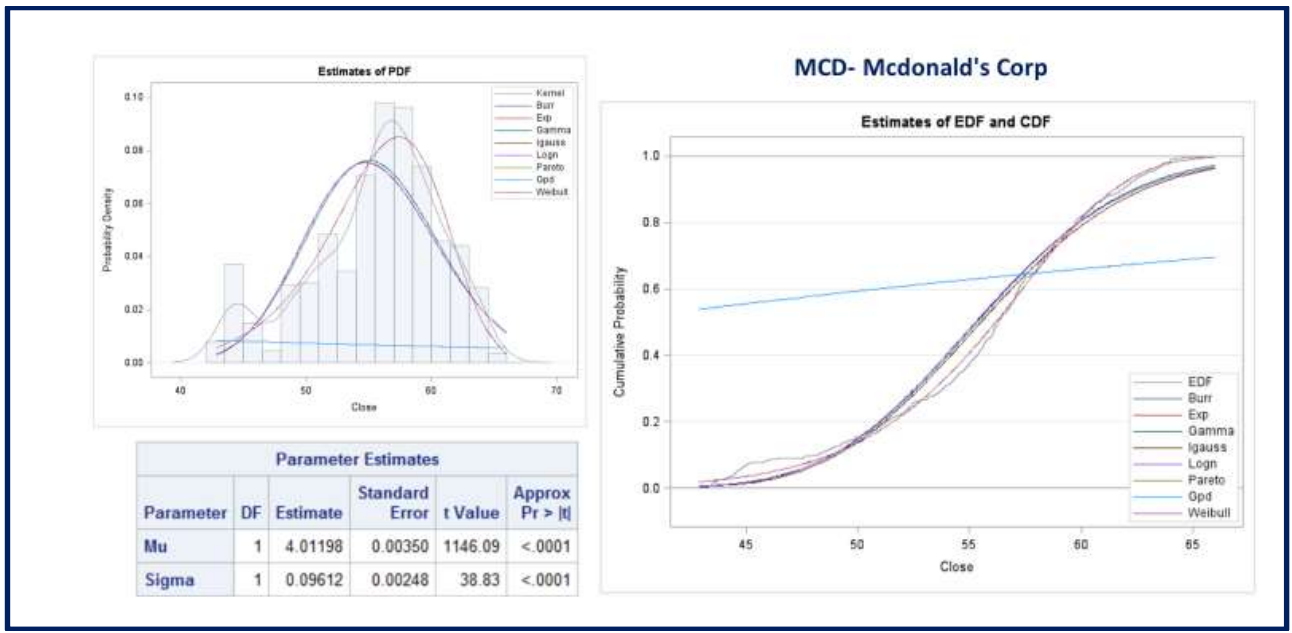

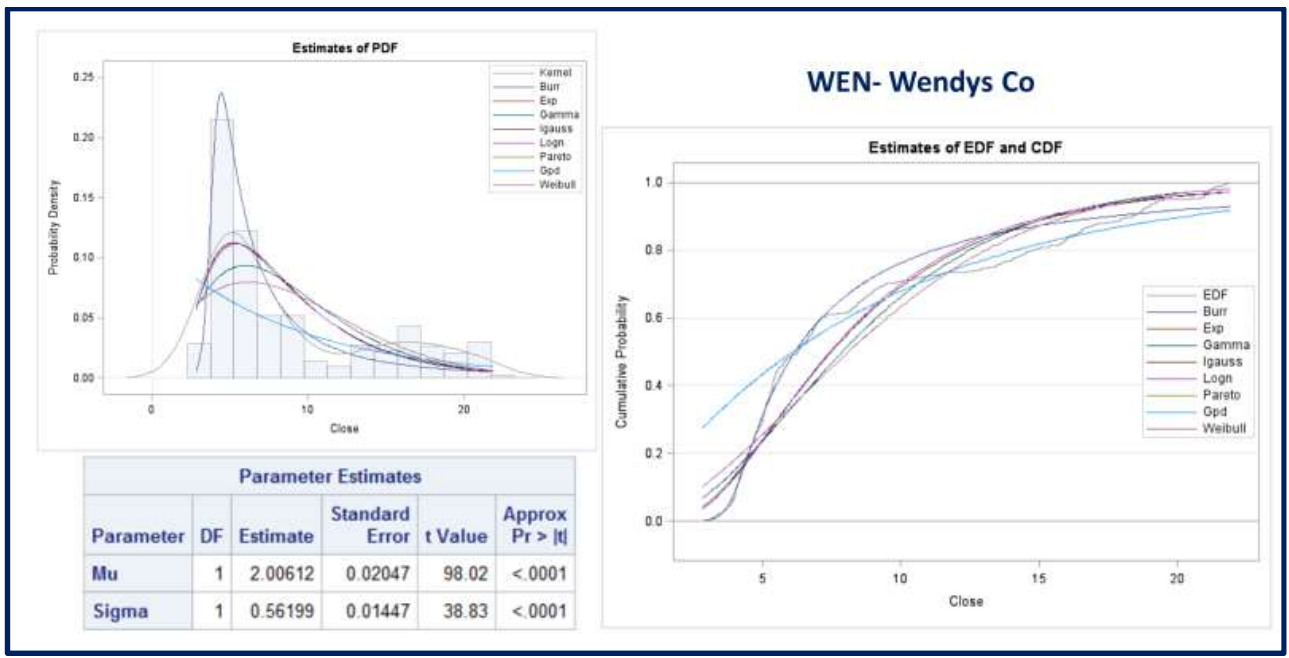

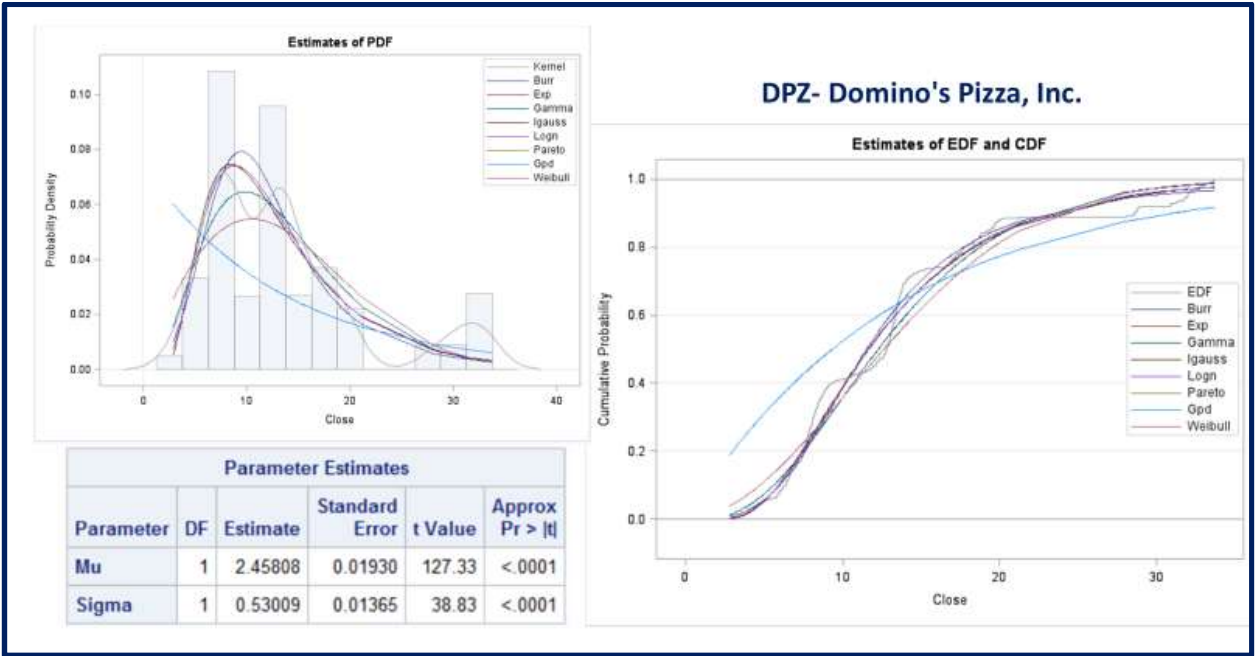

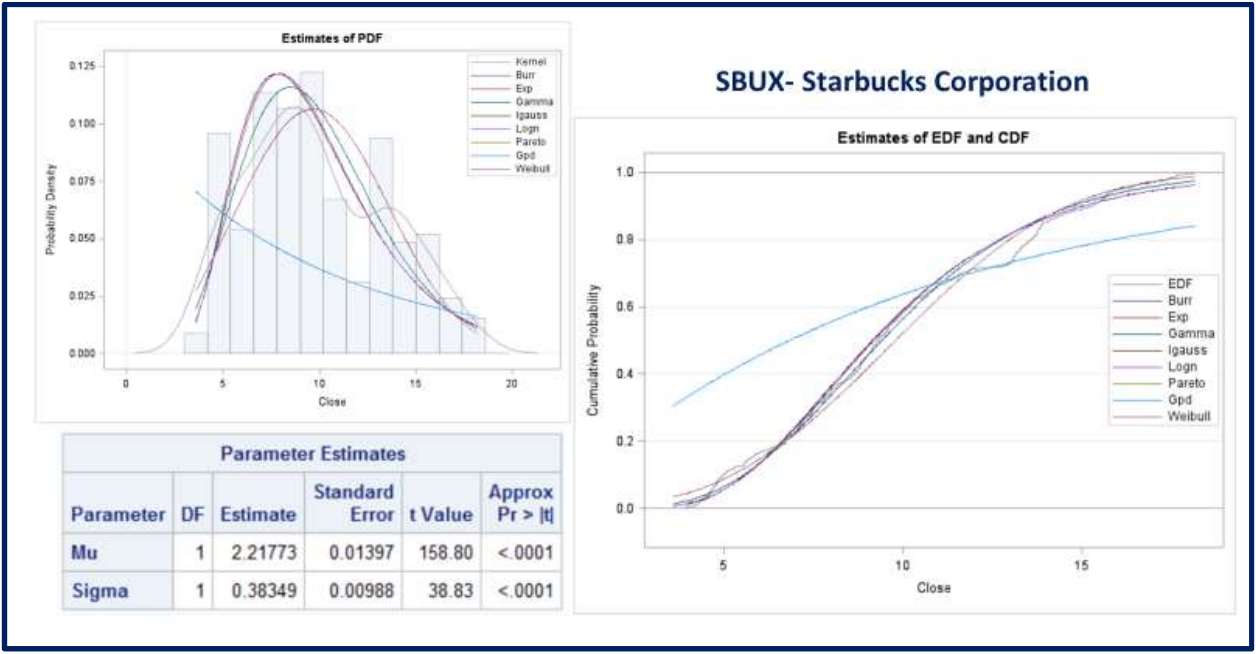

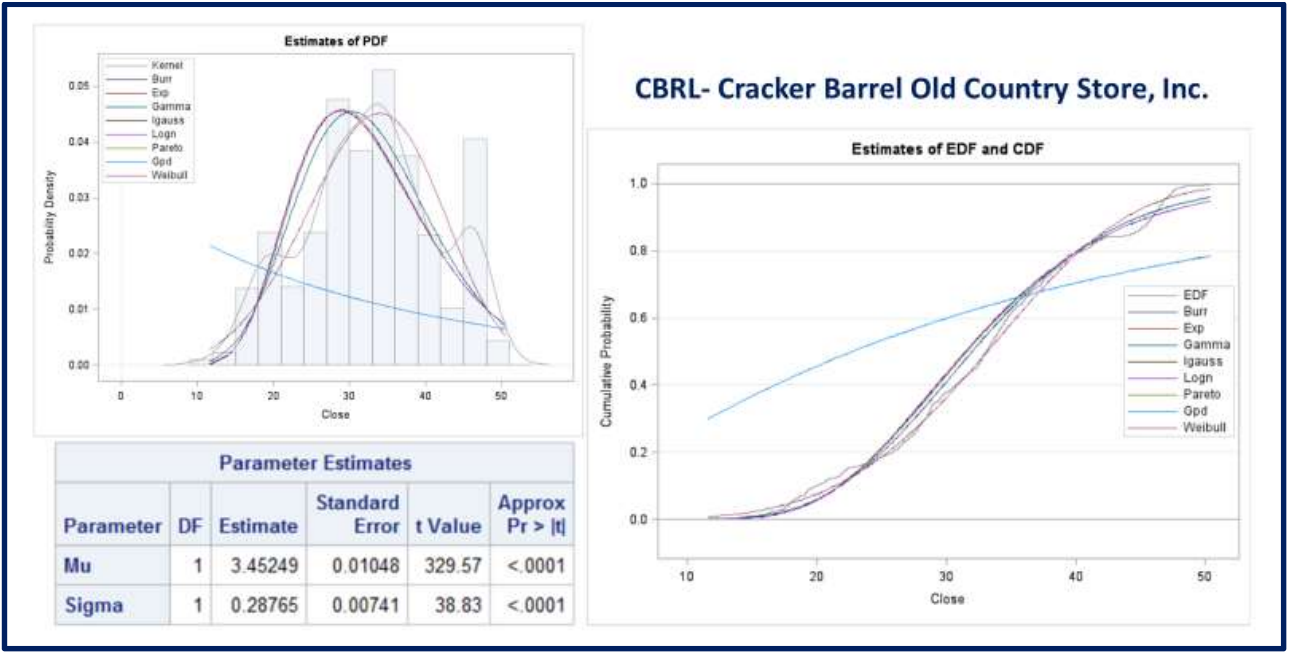

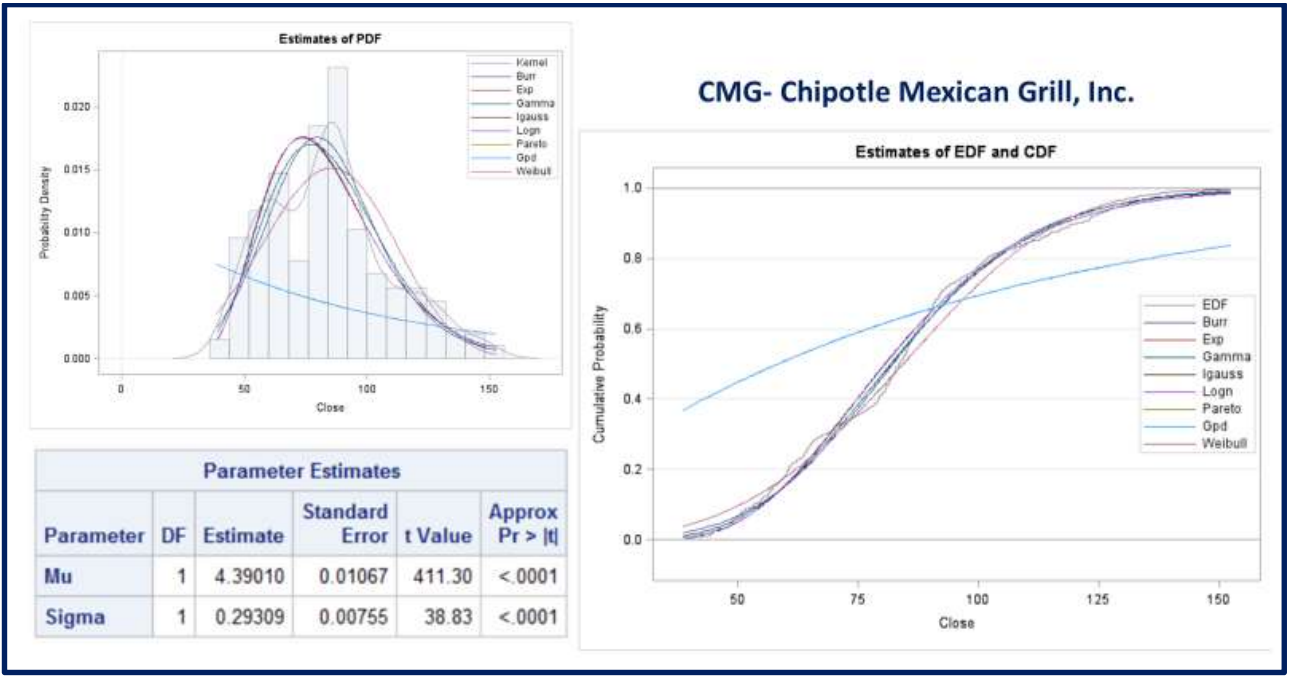

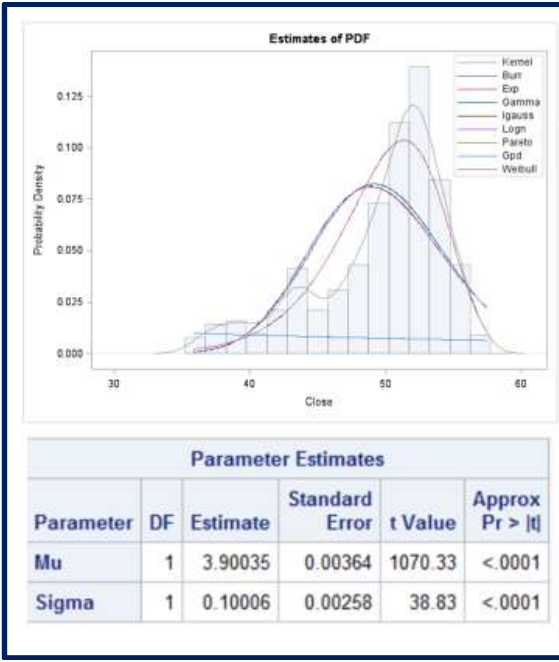
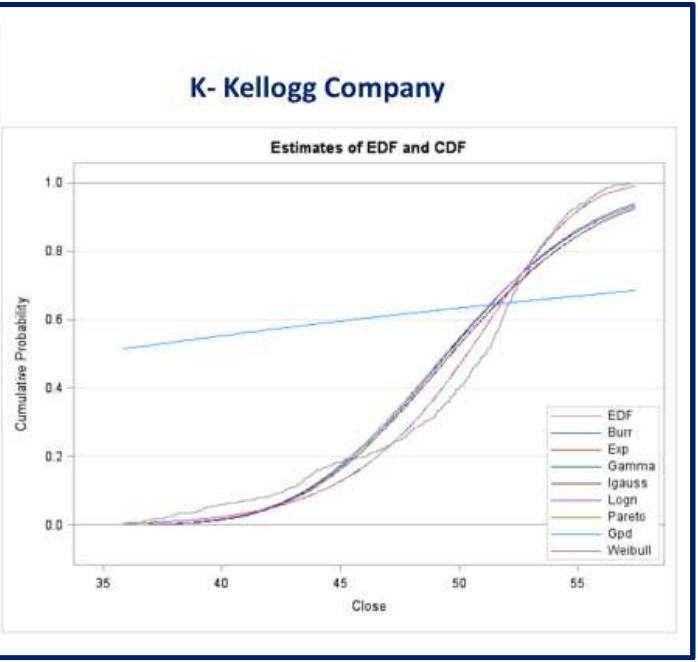

# K- Kellogg Company

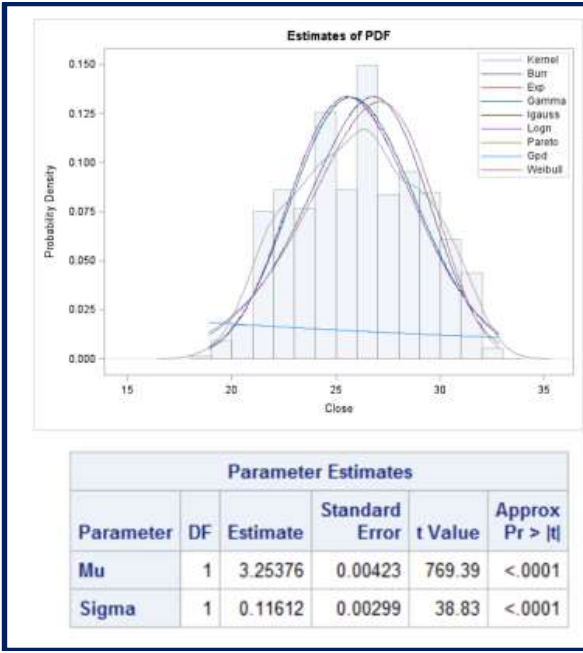
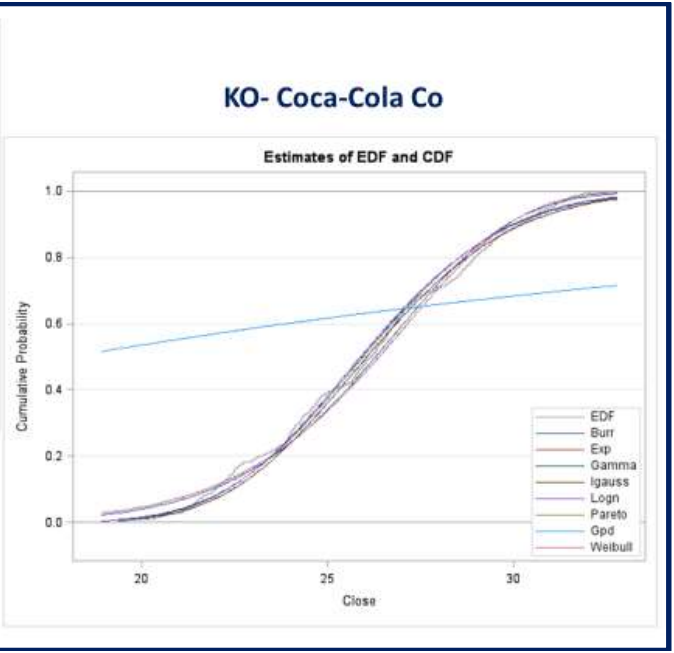

# KO- Coca-Cola Co

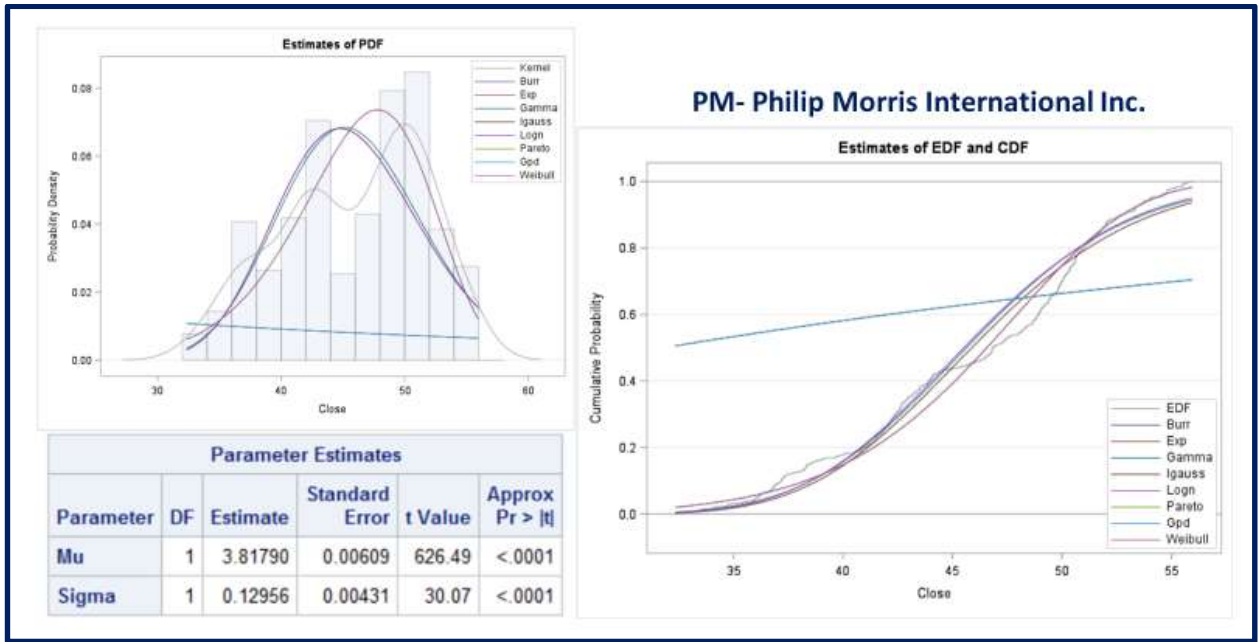

Conclusion: The franchised industry has a buffer layer between the parent franchisers and direct customers. This buffer layer creates a different distribution in the daily stock price even if the financials do not show a marked difference among these franchised and non-franchised companies. The franchised company like MCD shows the resilience and holds the price steady even in the turmoil of recession. Whereas WEN and DPZ fall below the sustainable level and cannot recover as other similar food related companies.

References:

Blair, R.D. and Lafontaine (2005), "The Economics of Franchising", Cambridge University Press

Bo, W., Dimitrova, Z.I., and Vitanov, N.K. (2019), "Statistical Analysis of the Water Level of Huang He River (Yellow River) in China", Jun 2019, arXiv:1906.00168v1 [stat.AP]

Combs, James G.; Michael, Steven C.; Castrogiovanni, Gary J. (2004), "Franchising: A Review and Avenues to Greater Theoretical Diversity", Journal of Management 30(6), pp. 907–931.

daCruz, J.P. and Lind, P.G. (2013), "The Bounds of Heavy-Tailed Return Distributions in Evolving Complex Networks," Jan 2013, arXiv:1109.2803v2 [q-fin.ST]

Dahle, Stephanie (2009), "McDonald's Loves Your Recession", Feb. 2009, Forbes

De Jonghe, Olivier and Vander Vennet, Rudi (2008), "Competition versus efficiency: What drives franchise values in European banking?", Journal of Banking & Finance 32, 1820–1835.

Dugan, Ann (1998), "Franchising 101", Association of Small Business Development Centers, edited by Ann Dugan, Upstart Publishing Company

Engle, Robert (2002) Dynamic Conditional Correlation, Journal of Business & Economic Statistics, 20:3, pp. 339-350,

Farfan, B. (2019), "All Publicly Traded U.S. Restaurant Chains", Retail Industry, Jan27, 2019, https://www.thebalancesmb.com/publicly-traded-us-restaurant-chains-2892798.

Felstead, A. (1993), "The Corporate Paradox- Power and Control in the Business Franchise", Routledge Publishers.

Frank, Nathaniel and Hesse, Heiko (2009). Financial Spillovers to Emerging Markets during the Global Financial Crisis, International Monetary Fund, 2009. ProQuest Ebook Central, working paper

Gim, Jaehee and Jang, SooCheong (Shawn) (2019), "Heterogeneous dividend behaviors: The role of restaurant franchising", International Journal of Hospitality Management 80, pp. 183–191.

Gim, Jaehee; Choib, Kyuwan; Janga, SooCheong (Shawn) (2019), "Do franchise firms manage their earnings more? Investigating the earnings management of restaurant firms", International Journal of Hospitality Management 79, pp. 70–77.

Gross, Daniel (2009), "Who won the Recession? McDonald's", 2009, MoneyBox, slate.com/business

Hsu, Li-Tzang (Jane) and Jang, SooCheong (Shawn) (2009), "Effects of restaurant franchising: Does an optimal franchise proportion exist?", International Journal of Hospitality Management 28, pp. 204–211.

Jennings, Lisa (2009), "On a roll: meals on wheels viewed as recession-proof franchise plan." Nation's Restaurant News 19 Oct. 2009: 4+. Business Insights: Essentials. Web. 15 Sept. 2019

Kizilersu, A., Kreer, M., Thomas, A.W., and Feindt, M. (2016), "Universal Behaviour in the Stock Market: Time Dynamics of the Electronic Orderbook," Physics Letters A 380 (2016) 2501-2512.

Macrotrends - The Premier Research Platform for Long Term Investors, https://www.macrotrends.net/

Martin, R.E. (1988), "Franchising and Risk Management", The American Economic Review, Vol. 78, No. 5 (Dec., 1988), pp. 954-968

Rhoua, Yinyoung; Lia, Yuan; Singalb, Manisha (2019), "Does managerial ownership influence franchising in restaurant companies?", International Journal of Hospitality Management 78, pp. 122–130.

Rocha, P., Raischel, F., daCruz, J.P., Lind, P.G. (2014), "Stochastic Evolution of Stock Market Volume-Price Distributions," Oct 2014, arXiv:1404.1730v2 [q-fin.ST]

Salar, Menekse and Salar, Orkide (2014), "Determining pros and cons of franchising by using swot analysis", 2nd World Conference on Design, Arts and Education DAE-2013, Procedia - Social and Behavioral Sciences 122, pp. 515 – 519.

Sayabaev, K. et al. (2016), "FINANCE, FRANCHISE AND THEIR IMPACT ON TOURISM", Journal of Internet Banking and Commerce, December 2016, vol. 21, no. 3

Solı́s-Rodrı́guez, Vanesa and Gonzá́lez-Dı́az, Manuel (2017), "Differences in contract design between successful and less successful franchises", Eur J Law Econ (2017) 44:483–502.

Sun, T. and Zhang, X. (2009). "Spillovers of the U.S. subprime financial turmoil to mainland china and Hong Kong SAR : Evidence from Stock Markets". International Monetary Fund, WP/09/166.

Wilson, Mark and Shailer, Greg (2015), "Information Asymmetry and Dual Distribution in Franchise Networks", Journal of Business Finance & Accounting, 42(9) & (10), pp. 1121–1153, November/December 2015

Xie, R. (2014), "Application of Modern Statistical Tools to Solving Contemporary Economic Problems: Evaluation f The Regional Agricultural Campaign Impact and the USDA Forecasting Efforts". PhD Dissertation, Clemson University